\documentclass[aps,prl,twocolumn]{revtex4}
\usepackage{graphicx}
\usepackage{natbib}

\begin{document}

\begin{widetext}

\title{Dynamics of the tuning process between singers.}

\author{R. Urteaga}
\affiliation{Centro At\'{o}mico Bariloche, Avda Bustillo km 9.5,
8400 San Carlos de Bariloche, R\'{\i}o Negro, Argentina.}

\author{P. G. Bolcatto}
\email{pbolcato@fiqus.unl.edu.ar}
\affiliation{Facultad de
Ingenier\'{\i}a Qu\'{\i}mica and Faculad de Humanidades y Ciencias,
Universidad Nacional del Litoral, Santiago del Estero 2829 3000
Santa Fe Argentina.}

\begin{abstract}
We present a dynamical model describing a predictable human behavior
like the tuning process between singers. The purpose, inspired in
physiological and behavioral grounds of human beings, is sensitive
to all Fourier spectrum of each sound emitted and it contemplates an
asymmetric coupling between individuals. We have recorded several
tuning exercises and we have confronted the experimental evidence
with the results of the model finding a very well agreement between
calculated and experimental sonograms.
\end{abstract}

\pacs{43.75.Cd, 43.75.Rs, 43.66.Ba}

\maketitle

\end{widetext}

\textit{Introduction}. When we think in music, we commonly do it in
relation to feelings and emotions arising from the sub-cortical
limbic system of the brain\cite {roederer04}. However, music or
sound perception is a very complex sequence of transductions,
beginning with the input of pressure waves to the ear and
ending with cognition operations developed in the brain's external neocortex%
\cite{guyton}. Consequently, an overall understanding of what music
means in human beings requires physical, biological, neural,
physiological and behavioral grounds\cite{roederer95}. In this
letter we are going to focus on the tuning process between singers.
The capability of human beings to sing in tune is strongly dependent
on his natural conditions, training and previous experience. Then,
results of tuning experiments can be very different even for the
same initial conditions. To avoid subjectivity we have restricted
the possible solutions by imparting a clear watchword oriented to
achieve tuning in the same note or in an octave. In this way we were
able to analyze experimentally basic human behavior and consequently
to propose a phenomenological mathematical model describing it.
While there are many works regarding
synchronization\cite{kuramoto,sakaguchi,sync1,sync2} (i. e.
\textit{phase} adjust) this is, up to our knowledge, the first model
that account for the evolution of spectra of frequencies interacting
between them.

A musical \textit{note} is a complex periodic oscillation that can
be discomposed into a sum of sinusoidal excitations, the
\textit{harmonics}, each one with a frequency multiple of a
particular frequency called the \textit{fundamental}. Then, if
$\omega _{0}$ is the fundamental frequency, the Fourier spectrum of
a note is composed by peaks at $\omega _{0}$, $2\omega _{0}$,
$3\omega _{0}$, etc. The \textit{pitch} indicates how high or low is
a particular note and is labeled with the value (or name) of the
fundamental. The relative intensities which each harmonic
participate in the sound define its
\textit{timbre.}\cite{roederer95}

A \textit{noise}, in turn, is a sum of excitations without any
relationship between the individual frequencies although the
boundary between music and noise is subjective and one can listen to
musicality in a given noise or find a noisy musical sound. In the
same way the idea of \textit{consonance} or \textit{dissonance} is
also a subjective, even cultural, concept. Nevertheless there are
physiological reasons to understand the consonance: the medium ear
contains a conduct with variable transversal section, the
\textit{cochlea}, inside which a stationary wave is formed. From the
hydrodynamical point of view, the cochlea is split-up in two
channels separated by the \textit{basilar membrane}. The differences
in pressure at both sides of the membrane produce deformations
resulting in a resonance pattern detected by a series of thin
receptors, the \textit{hair cells}, which are connected to neurons\cite{guyton}%
. Thus, the electrical signal sent to the brain is in fact a
transduction of the geometrical representation of the deformation of
the basilar membrane. The set of nodes of the stationary wave is
consistent with only one note (i.e., with only one Fourier spectrum)
and then, two notes will be more consonant as more nodes in common
they have.\cite{roederer04} Mathematically, the consonance is
reflected in a simple ratio between the
fundamental frequencies of each note. For example, if $\omega _{10}$ and $%
\omega _{20}$ are the fundamentals of two notes, a sequence from consonance
to dissonance is $\omega _{20}/\omega _{10}=1,2,3/2,4/3$ etc. The intervals
between $\omega _{10}$ and $\omega _{20}$ are denominated the same note,
octave, fifth, fourth, etc., respectively. For the purpose of this work, we
define tuning as the process in which two or more sound emitters change
their pitches in way of equaling all or part of their Fourier spectra.

\textit{Looking for the model}. In order to elaborate a mathematical
model which represents the main features of the tuning process, we
are going to extract basic ideas from some well-known responses of
the auditory system and also from prototype experiments:

(i) The interaction term have to be a function which goes to zero when the
ratio between frequencies is a simple fraction. In this way, we cover the
physiological and mathematical grounds of consonance.

(ii) Two complex tones with the same Fourier family but differing
only in that one of them has the fundamental missing, will be
listened by the singer as the same pitch. This ability of human
beings was characterized and explained through the concept of
virtual pitch perception introduced by E.
Terhardt\cite{terhardt2,terhardt}. We have verified this response
doing several experiments requesting to the singer to tune a
guitar's sound which was sequentially filtered in its lower
harmonics. In consequence, the functional response should be
proportional to all the spectrum more than a single frequency.

(iii) The point of subjectivity of ''how I listen to my partner and how
predisposed I am to interact with him'' can yields different final results
of tuning exercises even for the same couple of singers and with the same
initial condition. The model must contemplate this possibility.

\begin{figure}[ht]
\includegraphics[width=9cm]{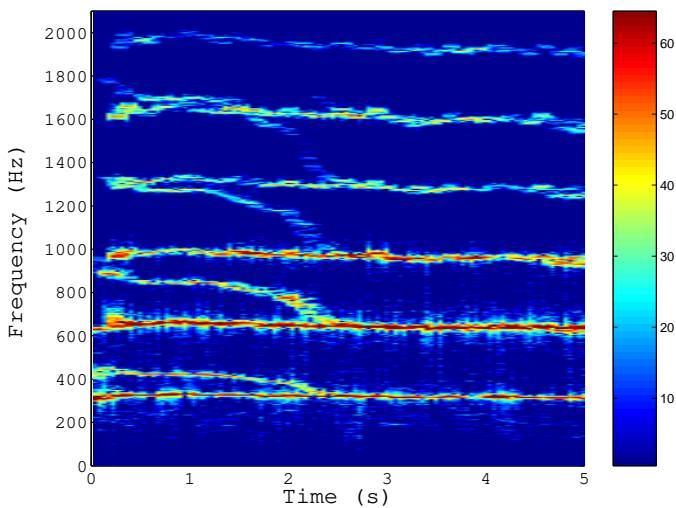}
\caption{Sonogram corresponding to a singer
tuning a note emitted and maintained by other singer.}
\label{Fig1}
\end{figure}

(iv) Finally, and with the aim to define terminology, it is useful
to analyze a very simple tuning exercise: a singer is asked to
maintain his pitch while the partner is moved until both are tuned.
None of them know the initial note of the other. The sonogram
(temporal evolution of the Fourier spectrum) of this exercise is
shown in Fig 1. At the beginning there is a brief interval of about
0.2 s in which the singers locate their initial note then, an period
of approximately one second lapses, and finally they effectively
start the exercise. We interpret the first interval as the necessary
time to accommodate the singing apparatus (vocal cords, resonators,
air emission, etc.) to produce a musical sound. The time spent in
this action can be reduced with training. The second stage is
necessary to perform the cognitive operation to listen to all the
notes emitted and to take the decision to move the pitch up or down
to tune. The remaining time is dedicated to feedback in order to
achieve tuning. We are going to name this last stage as
\textit{dynamical tuning}. In the example of Fig. \ref{Fig1}, and
because the watchword imparted, one of them acts like if he does not
listen to the other. In other words, there is an asymmetric coupling
between them.

Keeping in mind these precepts, we propose the following set of equations
describing the coupled dynamical evolution of complex tones emitted by $N$
singers:

\begin{equation}
\frac{d\omega _{i0}}{dt}=\frac{1}{N}\sum_{j}\sum_{\mu }K_{ij}I_{j\mu }\sin {%
(2m\pi }\frac{{\omega _{j\mu }}}{{\omega _{i0}}}{),}\text{ }i=1,...,N.
\end{equation}

$\omega _{j\mu }$ is the frequency of the $\mu ^{th}$ harmonic of the $%
j^{th} $ individual within the group, $I_{j\mu }$ is the relative intensity
of the corresponding harmonic in the Fourier expansion defining the timbre
of the sound, $K_{ij}$ is an off-diagonal matrix $(K_{ii}=0)$ representing
the effective relative magnitude of the coupling between pairs of singers
and $m$ is an integer constant.

The sine argument is the responsible to drive the stability of the equations
since when the relationship $m\omega _{j\mu }/\omega _{i0}$ is a integer,
the temporal derivative goes to zero. This sine function is indeed the key
point of the model. The condition for the roots works as the mathematical
representation of the natural behavior of singers to maximize the
coincidence of nodes in the resonance pattern of the basilar membrane. In
this sense, the goodness of the sine-like interaction is independent of $m$
since regardless the particular value of $m$, this functional response fits
the requirements of point (i). In a more general approach, the set of Eqs.
(1) should include a sum over $m$ but, as we are going to see later, the
trends of the experimental records can be reproduced with only one family of
$m-$like functions.

The absolute magnitude of the coupling is given by all the right side of Eq.
1 and the product $K\times I$ defines the temporal scale of the process. A
coupling proportional to all $I_{j\mu }$ guarantees a response to the
Fourier family more than a frequency in particular [point (ii)]. $K_{ij}$
can be thought in zero order approximation as a magnitude of the volume of
the emission, but as this matrix is asymmetric in general $(K_{ij}\neq
K_{ji})$, it can contemplate the alternative that one of the singers emits
always the same pitch independently of the movement of the rest. $K_{ij}=0$
means ''the $i$-singer is not coupled with the $j$-singer'' either because
he does not listen to the group or he has decided not to change his pitch.
By adopting different values for $K_{ij}$ we can obtain different final
results for the same pair of singers and starting with the same notes
respectively [point (iii)].

By construction, this model is oriented to describe the dynamical tuning, i.
e., when the singers start to move their frequencies by interaction.

\textit{Results and Discussion}. The individuals selected to all the
experiences were non-professional singers but most of them have or
had some training in collective singing. We formed 24 pairs of
singers and we recorded more than one hundred experiences. The
exercises were simple: firstly the initial note is indicated; each
singer listens to only his own note. Then, they simultaneously start
and change their pitches until to find tuning. The watchword was
\textit{''to arrive to the same note''} , which for a medium-trained
singer covers the possibility to tune in an octave. This last
alternative is more probable when the separation between the initial
pitches is large and/or when we treat with a female-male couple. We
have not taken into account those records in which the watchword was
not properly understood. We also discarded records in which one of
the singer is near to the limit of his range.

The numerical resolution of the system of equations was done through
an one step solver based on a Dormand-Prince-Runge-Kutta
formula\cite{rk} in which the frequencies were assumed constants in
the brief interval corresponding to the discretization adopted
($\simeq $ 1 ms). The numerical absolute error for the fundamental
frequencies was $10^{-5}$Hz. The initial values of fundamental
frequencies and harmonic intensities were extracted from an Fourier
analysis of a small initial interval of the {\it experimental}
sonogram. Strictly, the harmonic intensities change with the frequency $%
\left[ I=I\left( \omega \right) \right] $. However this $\omega -$dependence
is noticeable only when the pitch of the sound emitted is close to the
boundaries of the range (especially upper limit). Therefore, considering
that most of the exercises imply pitches in the medium region of the range,
we assume the harmonic intensities as constants $\left[ I\neq I\left( \omega
\right) \right] $.

\begin{figure}[ht]
\includegraphics[width=9cm]{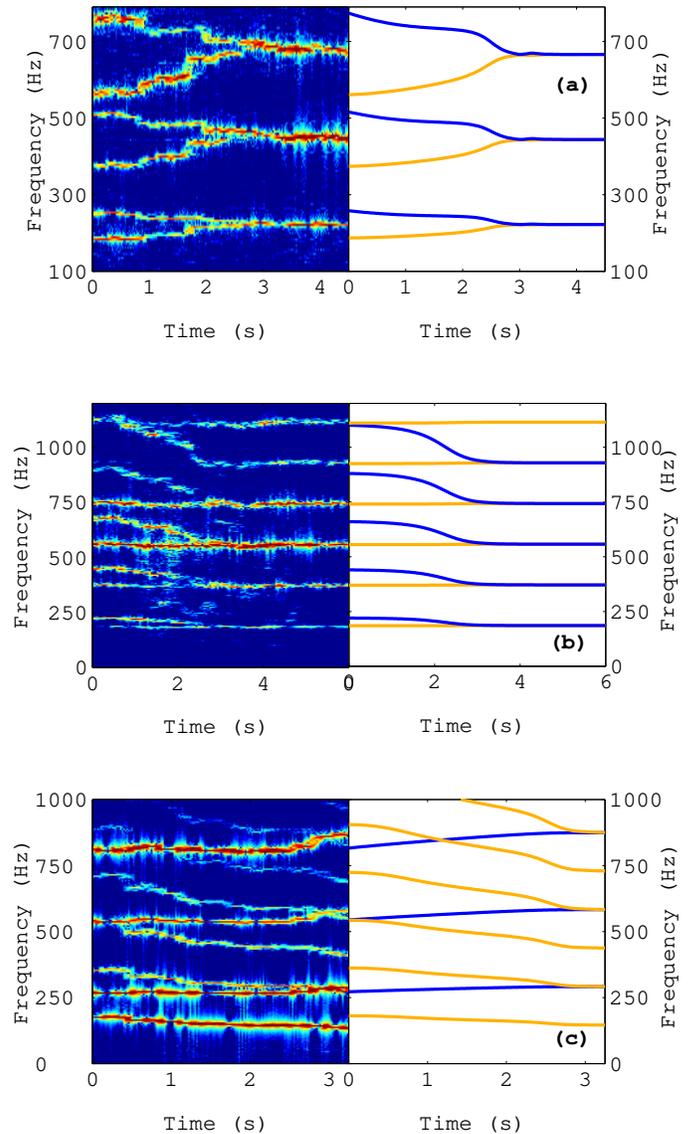}
\caption{Tuning at the same note. Left panel: Experimental sonogram
corresponding to the dynamical tuning. Same colormap as in Fig. 1.
Right panel: Results of the model given by Eq. (1) with $m=1$. The
value of the fundamental frequencies and the relative intensities of
the harmonics are taken from the Fourier Analysis of the interval 0s
- 0.5s. The drawing of simulations does not take into account the
Fourier intensities. (a) a baritone and a mezzo-soprano with
$K_{21}/K_{12}=1$. (b) The same as (a) for two tenors with
$K_{21}/K_{12}=25$. (c) The same as (a) for a baritone and a soprano
with $K_{21}/K_{12}=-1$.}
\label{Fig2}
\end{figure}

Figure \ref{Fig2} shows results for typical examples of tuning
exercises and its corresponding simulation. We have drawn on left
panel the experimental sonogram of the dynamical tuning and on right
panel the results of the model given by Eqs. 1. Fig. \ref{Fig2}a
corresponds to a baritone and a mezzo-soprano who start with
relatively near pitches ($\omega _{10}=187$ Hz and $\omega
_{20}=258$ Hz) and after 2.5 s they converge in a common note of
intermediate value ($\omega =219$ Hz). Both experimental and
theoretical results show an asymmetric dynamical evolution of each
spectrum in spite of the coupling is symmetric in this case
$(K_{21}/K_{12}=1)$. In Fig \ref{Fig2}b the exercise for two tenors
is drawn, one of them practically maintaining (by own decision) his
pitch. Here we can observe as the initial interval means a
type of consonance since there is a coincidence in harmonics of high order $%
(\omega _{20}/\omega _{10}=220$ Hz$/185$ Hz$\simeq 6/5)$ but as the
instruction is to move towards to the same note one of the singers changes
his pitch up to lock all the spectrum with the other. In this case we
reproduce the experimental evolution by adopting an asymmetric coupling $%
(K_{21}/K_{12}=25)$. Figure \ref{Fig2}c is an example of a tuning in
an octave for a baritone-soprano couple. The initial interval is a
fifth $(\omega _{20}/\omega _{10}=272$ Hz$/181$ Hz$\simeq 3/2)$ and
after the dynamical tuning they converge to an octave with
fundamental frequency of $\omega _{10}=146$ Hz for the baritone. In
this example the ratio between effective couplings is
$K_{21}/K_{12}=-1$. In all the cases the time required for the
dynamical tuning is about 1.5 s - 2 s independently of the training
or the quality of the singer.

\begin{figure}[t]
\includegraphics[width=8.5cm]{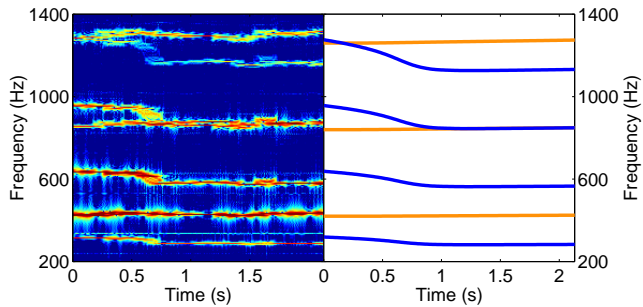}
\caption{A tenor and a mezzo-soprano tuning in consonance. Left
panel: Experimental sonogram. Same colormap as in Fig. 1. Right
panel: Results of the model given by Eq. (1) with $m=2$. The
fundamental frequencies and the relative intensities of the
harmonics are taken from the Fourier Analysis of the interval 0s -
0.2s. In this case $K_{21}/K_{12}=-0.15$.}
\label{Fig3}
\end{figure}

Theoretical results shown in Fig. \ref{Fig2} are very encouraging
since they reproduce almost exactly the experimental records but it
is worth to mention a word of caution. The minus sign in the
relation between effective couplings of the simulation presented in
Fig. \ref{Fig2}c seems to be non intuitive. However it is not a
conceptual barrier since in this case we need that the fundamental
frequencies go away one of another in order to tune in an octave.
So, the minus sign changes the direction of the derivative
facilitating the movement of fundamentals in the correct sense. We
remark
that we have wanted to fit the experimental records with only one type of $%
m- $like functions. In the context of this paper, the constant $m$
works as a degree of freedom of the model. In many cases -mainly
when there is not tuning at the same pitch- the model with $m=1$ is
not able to reproduce the experimental evidence although by fixing
$m=2$ we recover a good agreement. Clearly, the stability domains in
the time scale selected for the equation system change with $m$ and
then it is necessary to analyze what is the proper value of $m$ for
each case. This additional degree of freedom allow us to explore
other possible solutions. Figure 3 shows an interesting situation in
which we have changed the watchword asking to the singers
\textit{''to arrive to a pleasant sensation''. }Here we can study
what consonance means for each couple since the watchword can be
interpreted in a more subjective fashion. The example shown in Fig.
\ref{Fig3} is part of an exercise lasting 12 s approximately in
which the singers cross several stages of dynamical tuning. The
sequence was firstly a fourth and then three different fifths, each
one in a more comfortable sector of their ranges. We selected
the first movement from a fourth $(\omega _{20}/\omega _{10}=420$ Hz$/319$ Hz%
$\simeq 4/3)$ towards a fifth $(\omega _{20}/\omega _{10}=425$ Hz$/283$ Hz$%
\simeq 3/2)$. The record was reproduced by taking $K_{21}/K_{12}=-0.15$ and $%
m=2$. We notice that because his subjectivity the results emerging from this
second watchword were diverse and very singer-dependent and we not always
reached a good simulation.

\textit{Conclusions}. As a summary, in this work we propose a model
describing a particular and predictable human behavior like the
tuning process between singers. The calculations were done taking as
input parameters the \textit{experimental} values of initial
fundamental frequencies and harmonic intensities. We were able to
reproduce almost exactly the dynamical evolution for several
situations and we believe that this model containing the main
features of the tuning process could be the starting point to
further investigations in this field.

\textit{Acknowledgments}. The authors deeply acknowledge to Mario C.
G. Passeggi for the critical reading of the manuscript and to all
the persons that perform the experiences, in particular to the
members of the choir ''Vocal Consonante '', and the vocal group ''8
de canto'' and their musical directors, Abel Schaller and Mart\'{i}n
Sosa respectively. R. U. and P. G. B. are partially supported by
CONICET.

\smallskip

\end{document}